\documentclass[preprint2]{aastex}

\slugcomment{Not to appear in Nonlearned J., 45.}

\shorttitle{A Transmission-Filter Coronagraph: Design and Test}
\shortauthors{Ren et al.}

\begin{document}


\title{A Transmission-Filter Coronagraph: Design and Test}


\author{Deqing Ren\altaffilmark{1,2,3} , Jiangpei Dou\altaffilmark{1,2} and Yongtian Zhu\altaffilmark{1,2}}

%\affil{National Astronomical Observatories / Nanjing Institute of Astronomical Optics \& Technology, Chinese Academy of Sciences, Nanjing 210042, China.}
%
%\affil{ Key Laboratory of Astronomical Optics \& Technology, Nanjing Institute of Astronomical Optics \& Technology, Chinese Academy of Sciences, Nanjing 210042, China.}
%%\and
%\affil{  Physics \& Astronomy Department, California State University Northridge, 18111 Nordhoff Street, Northridge, California 91330-8268;\email{ren.deqing@csun.edu.}}


\altaffiltext{1}{National Astronomical Observatories / Nanjing Institute of Astronomical Optics \& Technology, Chinese Academy of Sciences, Nanjing 210042, China.}
\altaffiltext{2}{Key Laboratory of Astronomical Optics \& Technology, Nanjing Institute of Astronomical Optics \& Technology, Chinese Academy of Sciences, Nanjing 210042, China.}
\altaffiltext{3}{Physics \& Astronomy Department, California State University Northridge, 18111 Nordhoff Street, Northridge, California 91330-8268;ren.deqing@csun.edu.}


\begin{abstract}
We propose a transmission-filter coronagraph for direct imaging of Jupiter-like exoplanets with ground-based telescopes. The coronagraph is based on a transmission filter that consists of finite number of transmission steps. A discrete optimization algorithm is proposed for the design of the transmission filter that is optimized for ground-based telescopes with central obstructions and spider structures. We discussed the algorithm that is applied for our coronagraph design. To demonstrate the performance of the coronagraph, a filter was manufactured and laboratory tests were conducted. The test results show that the coronagraph can achieve a high contrast of $10^{-6.5}$ at an inner working angle of $5\lambda/D$, which indicates that our coronagraph can be immediately used for the direct imaging of Jupiter-like exoplanets with ground-based telescopes.
\end{abstract}



\section{Introduction}
With over 400 extra-solar planets detected today, most via the indirect detection techniques such as the radial velocity approach, the direct imaging of exoplanets is receiving increasing attention. Direct detection of photons from exoplanets will allow us eventually to achieve the most critical scientific goals in the astrophysics such as searching for another Earth. For Jupiter-like exoplanets, a moderate contrast on the order of 10$^{-6}$ is required for the direct imaging (Marley et al. 2007, Marois et al. 2008), which can be done on a ground-based telescope.

In recent years, many coronagraphs have been proposed which can theoretically reach a high contrast on the order of $10^{-10}$ with inner working angle (IWA) of few $\lambda/D$ (Kasdin et al., 2003, Vanderbei et al. 2004, Guyon et al. 2006, Ren \& Zhu 2007, Enya et al. 2008). However, most previous coronagraphs were optimized for dedicated off-axis telescopes that have no central obstructions and spider support structures, which is not suitable for today's large ground-based telescopes. The existence of central obstruction and spider structure will introduce further diffraction, which makes the design of a high-contrast coronagraph difficult. Recently, Soummer et al. have discussed a coronagraph that uses transmission apodized pupil (Soummer et al. 2009), in which they use an analytic function called generalized prolate spheroidal function to design the apodized pupil.

Although the transmission apodized pupil realized by metallic coating is one of the promising techniques, it is intrinsically chromatic and may induce wavefront phase errors by a metal layer of variable thickness. To overcome these problems, a technique called microdot was proposed recently (Martinez et al. 2009a, Martinez et al. 2009b). The microdot technique is more complex in the design and manufacturing. We will show here that by carefully choosing coating material and using optimization algorithm our transmission filter that is based on the metallic coating technique can also deliver a similar or slightly better performance, but at a much low cost.

In this work, we report our recent development for the design and laboratory test regarding a transmission-filter coronagraph. Our design is based on a discrete optimization algorithm, in which only finite number of transmission steps/pixels is used. Such a discrete optimization approach is especially suitable for telescopes with specific central obstructions and spider structures. We future include the phase error in the optimization, which makes it more realistic for the real situation. Wideband imaging is also discussed. In Section 2, we describe our discrete optimization algorithm. The laboratory test of the coronagraph is discussed in Section 3. Conclusions are presented in Section 4.

\section{Coronagraph Design}
The general idea of using transmission filters with finite number of transmission steps for high-contrast imaging was discussed by Ren \& Zhu (2007). Here, we discuss the algorithm that uses numerical approach to find the optimization solution for a specific situation with telescopes that have different obstructions and spider structures. Our coronagraph uses finite number of transmission steps where the transmission is identical in each step. The transmission filter is located on a conjugated pupil image plane, where the light is collimated. Star and exoplanet images are formed on the focal plane of the coronagraph. Our transmission filter is realized by metallic coating material deposited on a glass substrate, in which the transmission is controlled by the adjustment of the thickness of the coating material. Since the transmission of the filter is variable as a function of the radius, the optical path is not identical in each step of the filter, which will introduce a phase error. Assume the filter is circularly symmetrical around the center, and it is convenient to use a polar coordinate system. The point spread function (PSF) of the starlight on the focal plane ( $r$ radial coordinate) is related with the transmission filter ($r^{\prime}$ radial coordinate) by a the Fourier transform function as,
\begin{equation}
I(r)=|\vec{F}[A(r^{\prime})e^{-i\phi(r^{\prime})}]|^{2},
\end{equation}
where $\vec{F}$ represents the operation of the 2-dimensional Fourier transform. $A(r^{\prime})$ is the so called pupil function, which is determined by the transmission filter. $r$ and $r^{\prime}$ are the radii on the focal and filter planes, respectively. If the intensity is uniform on the pupil, the pupil function is simply the electric field of the transmission filter. $\phi(r^{\prime})$ is the possible phase error that may be introduced by the thickness variation of the filter coating material.

For a metallic coating material, its refractive is a complex number and can be expressed as $\tilde{n}=n+ik$, where $n$ is the refractive index indicating the phase velocity and $k$ is called the extinction coefficient, which indicates the amount of absorption when the electromagnetic wave propagates through the metallic material. The transmission of the metallic film is decreased with the thickness $d$ as (Born and Wolf 1999)
\begin{equation}
T=T_0e^{-4\pi kd/\lambda_0},
\end{equation}
where, $\lambda_0$ is the wavelength in the vacuum. $4\pi k/\lambda_0$ is called the absorption coefficient. The pupil function is related with the transmission as $A=\sqrt{T}$. By adjusting the thickness, one can change the transmission. The variation of the thickness will, however, induce a phase error, which will greatly degrade the performance if such a phase error is not considered in the design of a transmission filter. The phase error induced by the thickness difference is calculated as $\phi(r^{\prime})=2\pi(n-1)d/\lambda_0$. It is clear that for a fixed $n$ and transmission, a large $k$ will result in a small optical path difference $d$ and phase error. In addition, for the same $k$, the phase error will decrease at a longer wavelength.

The contrast is defined as the ratio of intensity on a specific location to the peak intensity on the PSF center, and is given as
\begin{equation}
C(r)\equiv I(r)/I(0).
\end{equation}
In the discovery area that is defined by the inner working angular distance and the outer working angular distance (OWA), assume the target contrast $C_t$ is a constant (such as $10^{-6.5}$), the algorithm that is based on the discrete optimization is to minimize the following equation
\begin{equation}
min ~~~~~~\{\sum_{IWA}^{OWA}|C(r)-C_{t}|\}.
\end{equation}
The algorithm is to optimize the contrast on a focal plane discovery area that is defined by the IWA and OWA. To get a good optimization result, a trade-off is needed among the target contrast, discovery area and transmission. For example, an over-low contrast may not be able to achieve and which may also result in a low transmission. Therefore, the design of the metallic transmission filter is to find the best transmission profile that satisfies equation 4. The discrete optimization algorithm has the advantage to be able to find an optimized solution for telescopes with specific obstructions and spider structures, and the step number used for the discrete optimization can exactly match the actual pixel number that are determined by the manufacturing spatial resolution. For example, in our filter design we use 50 steps, since the filter was made by Reynard Corporation who can make the filter with a spatial resolution of 50 pixels along a 15-mm radius of the clear aperture. In general, increasing the step number can increase the OWA, which was discussed on our previous work (Ren \& Zhu 2007).

Our discrete optimization includes the phase error that is induced by the thickness variation of the coating material. The phase error is not an independent parameter. It is associated with the thickness $d$ of the coating material and can be solved directly from equation (2). A large $k$ and a small $n$ will result in a small phase error. By carefully choosing the coating material, good performance with a contrast up to $10^{-6}\sim10^{-7}$, which is enough for the direct imaging of young Jupiter-like exoplanets with a ground-based telescope, can be achieved.

As a demonstration of the filter-transmission coronagraph, we present a design example. We choose Inconel as the metallic coating material that is widely used for the neutral metallic density filter. According to the data sheet provided by the Reynard Corporation, it has a complex refractive index $\tilde{n}=2.18+i3.05$ at 600nm wavelength, while the refractive index is $\tilde{n}=3.62+i6.54$ at 2000nm. Complex refractive indices at other wavelengths can be interpolated from the discrete data provided by the company. Since the variation of the complex refractive index, both phase error and transmission will slightly change at other wavelengths. The transmission filter is optimized at the design/optimized wavelength that is the central wavelength for a wideband imaging. Figure 1 shows the contrast at the 1.65 $\mu$m ($H$ band) optimized wavelength, which is designed with our discrete optimization algorithm. The contrast at the 1.825 $\mu$m non-optimized wavelength that is the end wavelength of the $H$ band is also calculated, which includes the variation of the phase and transmission because of the wavelength shift. For the non-optimized longer wavelength, the contrast is slightly improved at smaller angular distances while it is slightly degraded at larger angular distances. The contrasts at both wavelengths are better than $10^{-6}$ at an angular distance equal or larger than $5\lambda/D$, and the contrast difference between the optimized and non-optimized wavelengths is less than $10^{-0.5}$ in general. It is clear that the Inconel can be used for the high-contrast imaging over a good wavelength range.
%% In this section, we use  the \subsection command to set off
%% a subsection.  \footnote is used to insert a footnote to the text.


%% This section also includes several of the displayed math environments
%% mentioned in the Author Guide.

\section{Laboratory Test}
To demonstrate the performance of the transmission-filter coronagraph, one filter is designed. The metallic material of Inconel is deposited on a BK7 substrate and the transmission is controlled by the adjustment of thickness of the Inconel. The metallic coating filter, which consists of 50 steps along the aperture radius, was manufactured by Reynard Corporation. The filter has a clear diameter of 30mm with a central circular opaque region of 3.6-mm diameter, which corresponds to a linear obstruction of 12$\%$ . The width of the spider is 0.45mm, which takes 1.5$\%$ diameter of the clear aperture. The transmission error of the coating in manufacturing is less than 5$\%$ . For test purpose and measurement convenience, the filter is designed at the 632.8 nm Helium-Neon laser test wavelength. The overall throughput of the filter is ~31$\%$ . The filter and spider structure were manufactured individually, as shown in Figure 2.

The coronagraph optics consists of two transmission lenses. One is served as collimator while the other is used as camera lens that form a focal plane image of the test light source where a Starlight Xpress CCD detector array is used to measure the PSF. A spatial pinhole is used to create a perfect point light source. The transmission filter is located immediately after the collimator lens. We found that the multi-reflection from these lens curvature surfaces, the CCD detector glass window as well as the optical imperfect such as the dust particles introduces some scattered lights.  Nevertheless, the test shows that the coronagraph is able to deliver a contrast up to $10^{-6.5}$ at an IWA of $5\lambda/D$, which is consistent with our theoretical estimation. The PSF images with different exposure times are shown in Figure 3. In order to see the details of the low intensity areas on the PSF plane, the center and right panels in Figure 3 are overexposed. The strong bright vertical patterns in these two panels are due to the CCD image bloom.  Figure 4 shows the associated contrast along the PSF diagonal direction, in which the test contrast is shown in solid line, while the theoretical PSF profile is shown in dotted line. Compared with the theoretical profile, the test PSF has a slight deviation, which is introduced by the filter transmission error as well as possible residual wave-front error of the test lenses. However, such a deviation is well controlled and is at an acceptable level.

The precision of the transmission is always a concern for a transmission filter, which determines the performance of the coronagraph. The image of the illuminated transmission filter is recorded on the CCD detector array by using a replay optics that creates a filter image onto the CCD focal plane. The measured intensity distribution is compared with the design values. Figure5 (Left) shows the image of the intensity distribution of the test filter, in which small bulbs resulted from defect glass surfaces can be seen clearly. The comparison of the transmission section plot of the test filter and the design profile is shown Figure 6. It is clear that the test transmission curve agrees well with the design profile, except at the areas around the 2 intensity peaks, which introduces some deviation between the test and the theoretical PSFs as shown in Figure 4.

\section{Conclusions}
We have demonstrated how to design a transmission-filter coronagraph for wideband high-contrast imaging. Our design is based on the discrete optimization algorithm which includes the possible phase error that is induced by the thickness variation of the metallic coating material. The discrete optimization approach uses finite number of steps/pixels which is suitable for specific telescopes that have different sizes for the central obstruction and spider structure. Since phase error is also included in the discrete optimization, good agreements between the test and theoretical estimation are achieved. The coronagraph laboratory test has achieved a contrast of $10^{-6.5}$ at an angular distance of $5\lambda/D$ or larger, without any wave-front correction by using a deformable mirror. The design and test results indicate that our transmission-filter coronagraph can be used immediately for the direct imaging of hot Jupiter-like exoplanets with a ground-based telescope that is equipped with an adaptive optics system that can effectively correct the atmospheric turbulence.

\clearpage

\begin{figure}
\epsscale{.80}
\plotone{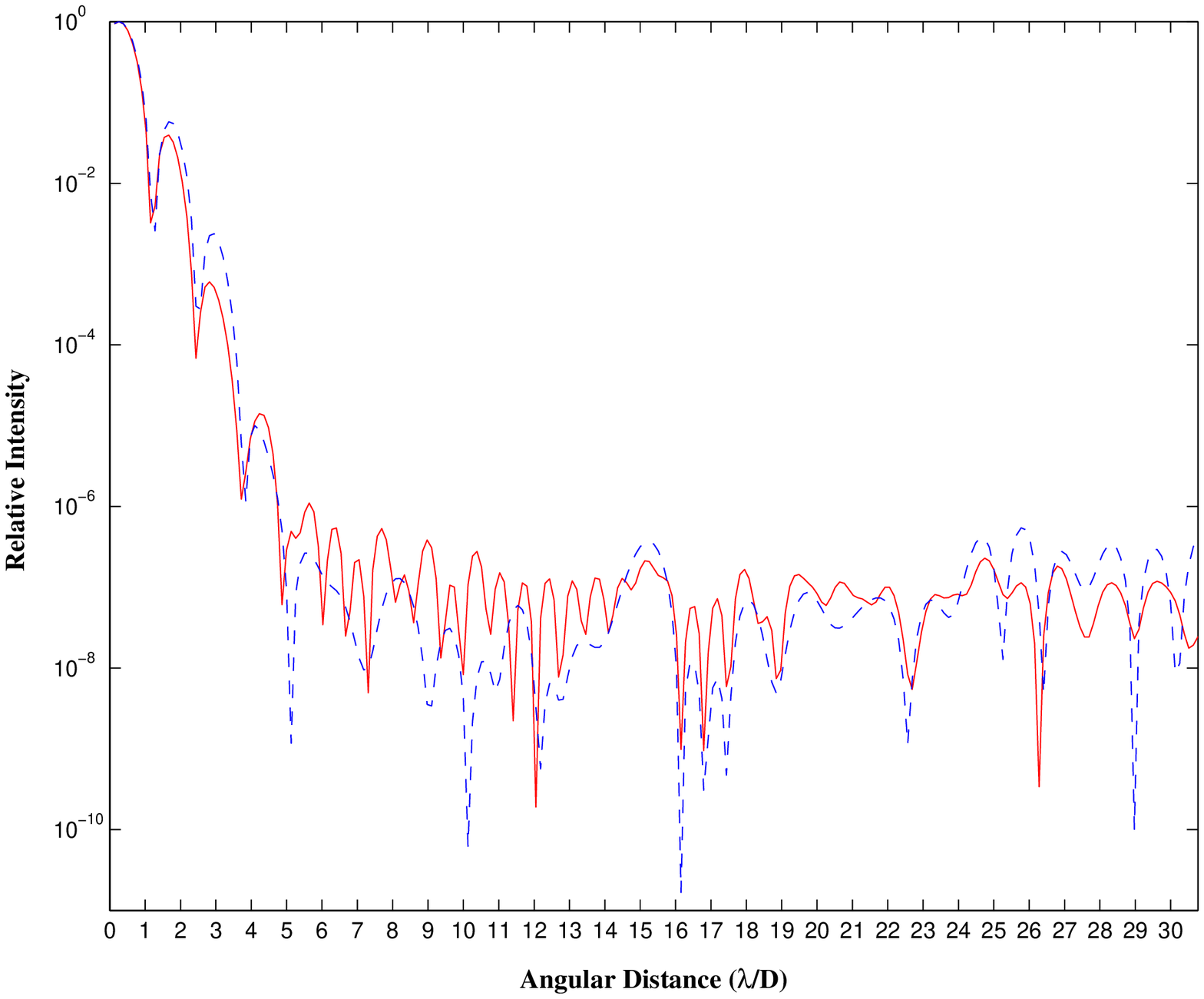}
\caption{The contrast at the 1.65 $\mu$m optimized wavelength (red solid line) and the 1.825 $\mu$m non-optimized wavelength (blue dotted-line).\label{fig1}}
\end{figure}

%% Here we use \plottwo to present two versions of the same figure,
%% one in black and white for print the other in RGB color
%% for online presentation. Note that the caption indicates
%% that a color version of the figure will be available online.
%%

\begin{figure}
\plottwo{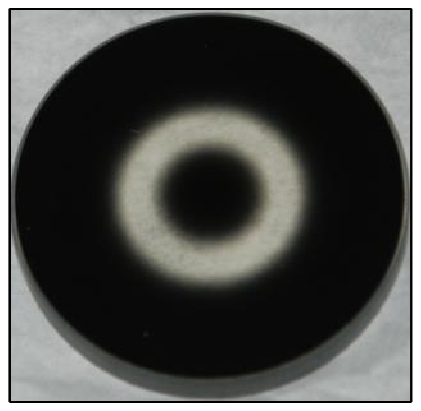}{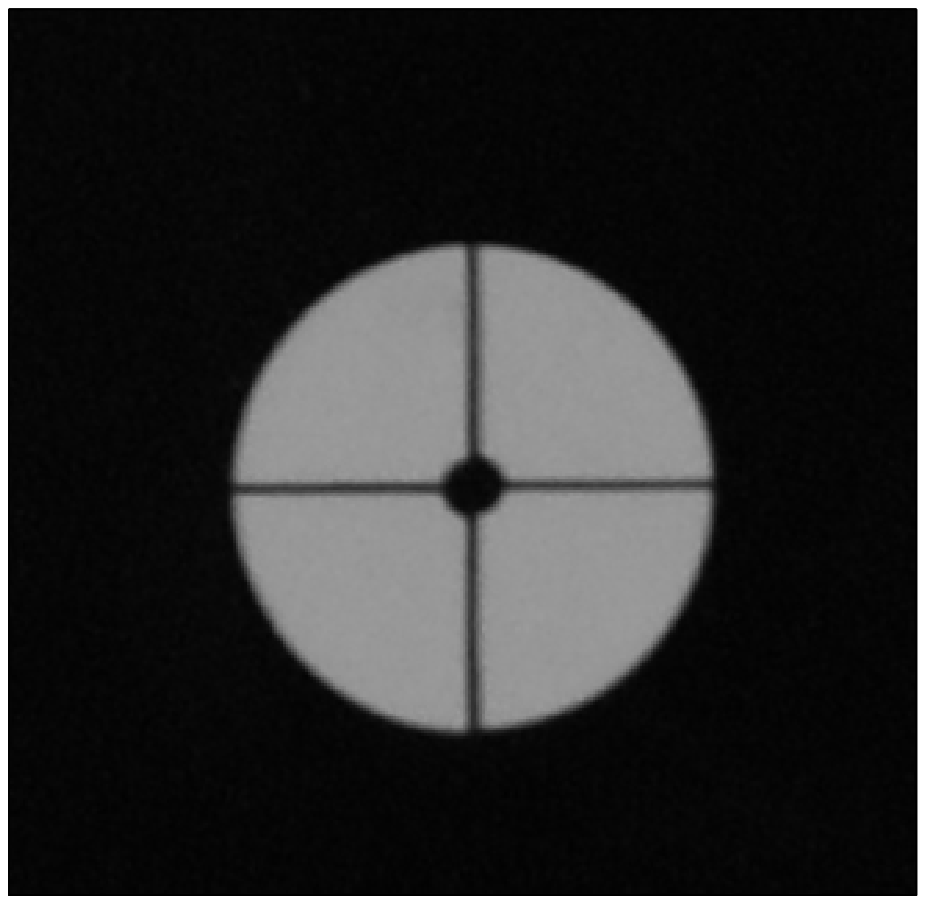}
\caption{Photograph of the finished filter (Left) and the associated spider structure (Right).\label{fig2}}
\end{figure}

\clearpage
%% This figure uses \includegraphics to scale and rotate the still frame
%\begin{figure}
%\plotthree{f3a.eps}{f3b.eps}{f3c.eps}
%\caption{PSF images under exposure times of 0.09s, 7.2s and 900s (from the left to the right panels), respectively.\label{fig3}}
%\end{figure}

\begin{figure}

\includegraphics[width=47mm]{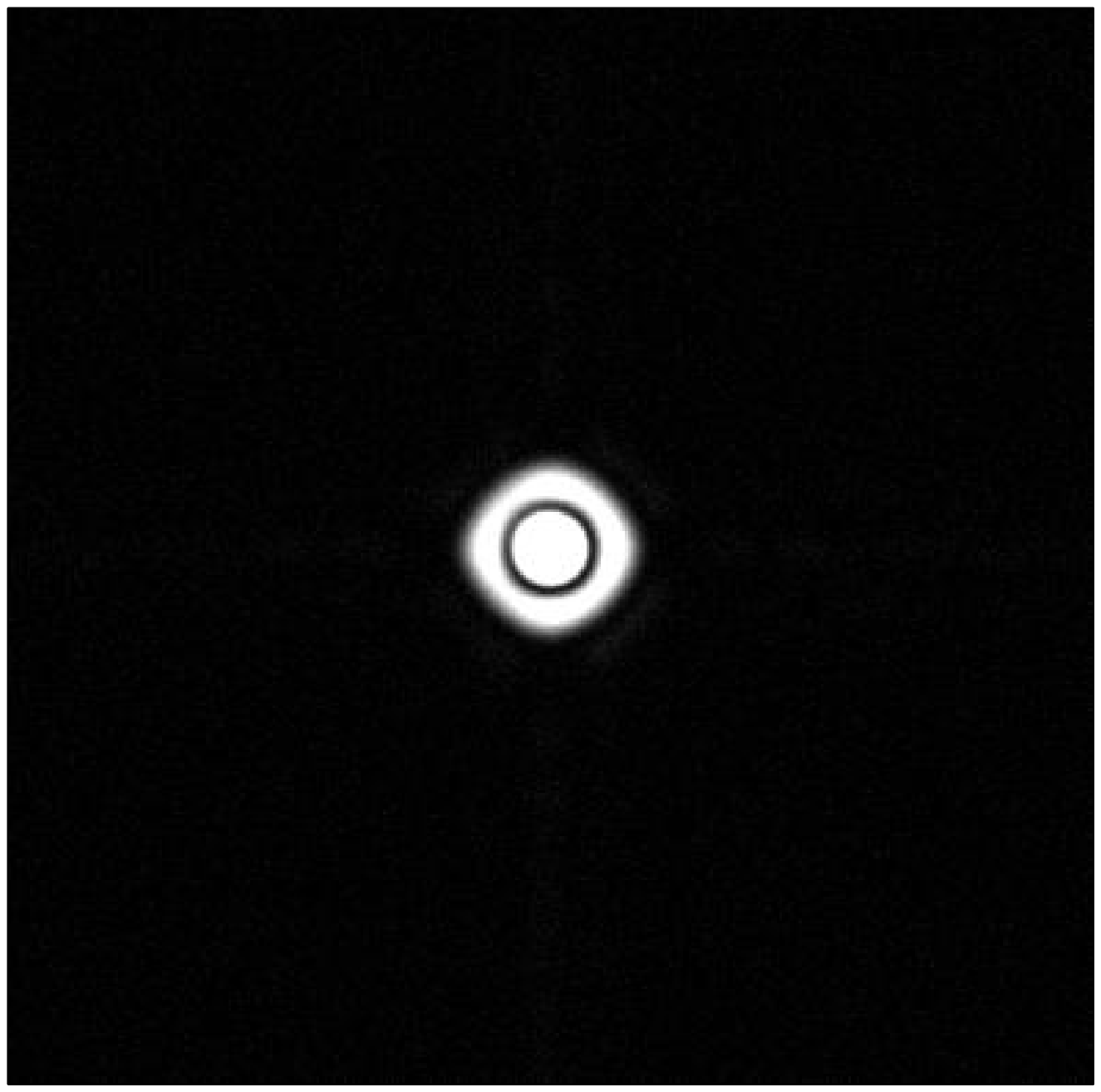}
\includegraphics[width=45mm]{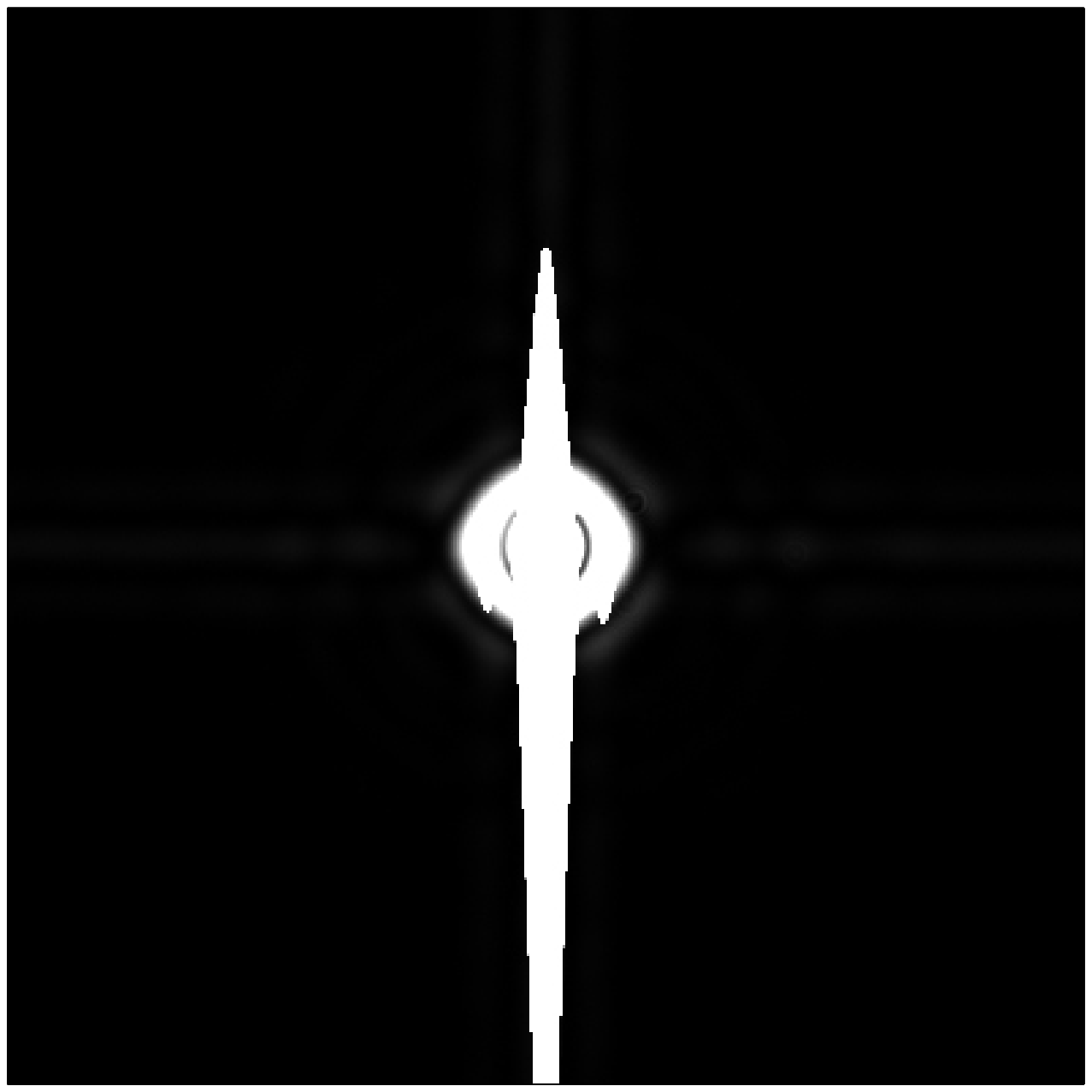}
\includegraphics[width=45mm]{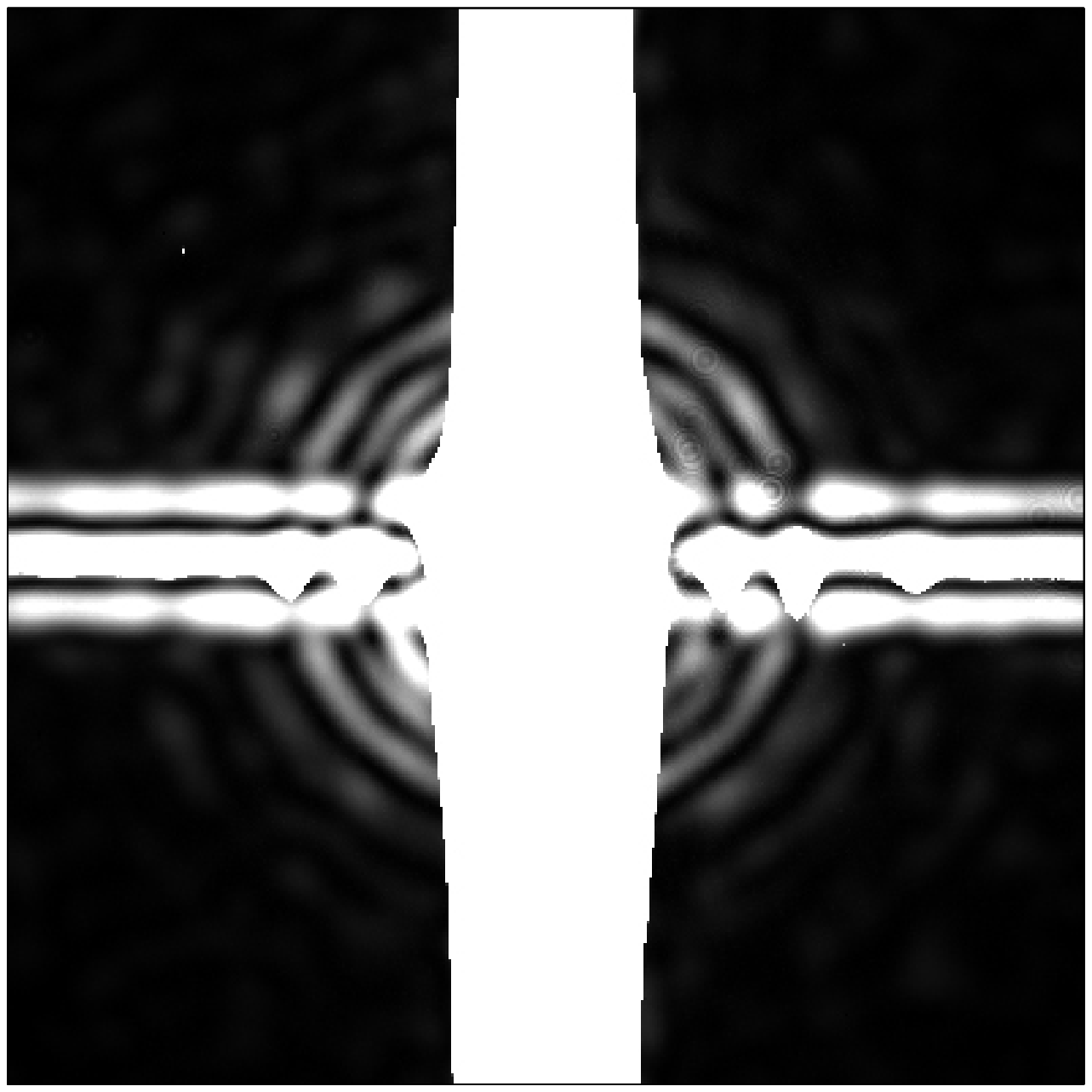}

\caption{PSF images under exposure times of 0.09s, 7.2s and 900s (from the left to the right panels), respectively.\label{fig3}}
\end{figure}

\begin{figure}
\epsscale{.80}
\plotone{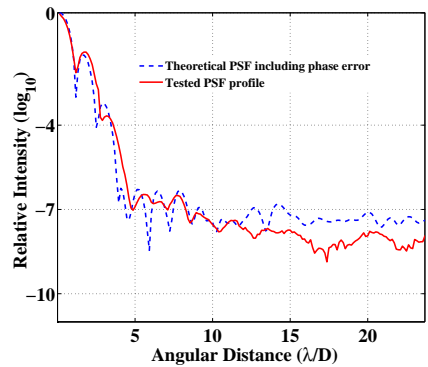}
\caption{Test (red solid line) and theoretical (blue dotted line) contrasts along the PSF diagonal direction. The test contrast achieves $10^{-6.5}$ at $5\lambda$/D or larger angular distance.\label{fig4}}
\end{figure}

\clearpage

\begin{figure}
\plotone{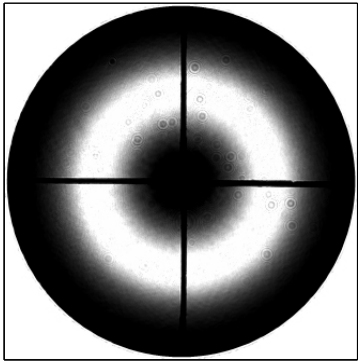}
\caption{Intensity image of the transmission filter. \label{fig5}}
\end{figure}

\begin{figure}
\plotone{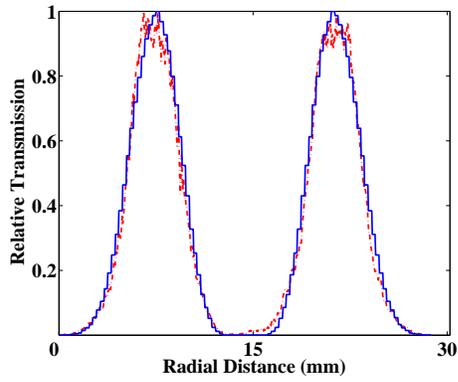}
\caption{The comparison of the design (blue solid line) and test (red dotted line) transmission profile.\label{fig6}}
\end{figure}
%% If you are not including electonic art with your submission, you may
%% mark up your captions using the \figcaption command. See the
%% No more than seven \figcaption commands are allowed per page,
%% so if you have more than seven captions, insert a \clearpage

%% Tables should be submitted one per page, so put a \clearpage before

%% Two options are available to the author for producing tables:  the
%% deluxetable environment provided by the AASTeX package or the LaTeX
%% table environment.  Use of deluxetable is preferred.
%%

%% Three table samples follow, two marked up in the deluxetable environment,
%% one marked up as a LaTeX table.

%% In this first example, note that the \tabletypesize{}
%% command has been used to reduce the font size of the table.
%% We also use the \rotate command to rotate the table to
%% landscape orientation since it is very wide even at the
%% Note also that the \label command needs to be placed

%% This table also includes a table comment indicating that the full
%% version will be available in machine-readable format in the electronic

%% You can append references to a table using the \tablerefs command.
\tablerefs{
(1) Born\& Wolf 1999; (2) Enya et al. 2008; (3) Guyon et al. 2006;
(4) Kasdin et al. 2003; (5) Marley et al. 2007: (6) Marois et al. 2008;(7) Martinez et al. 2009a;
(8) Martinez et al. 2009b;(9) Ren \& Zhu 2007; (10)Soummer et al. 2009; (11) Vanderbei, Kasdin, \& Spergel 2004.}

\end{document}